\documentclass[12pt]{article}
\pdfoutput=1
\usepackage{amssymb,amsfonts}
\usepackage{graphics,psboxit,amsmath}
\usepackage{subfigure}
\usepackage{graphicx}
\usepackage{verbatim}
\usepackage{color}
\usepackage{dsfont}

 \newcommand{\sss}[1]{{\scriptscriptstyle #1}}

\newcommand{\opc}[1]{c^{\phantom{\dagger}}_{#1}}
\newcommand{\opcd}[1]{c^{\dagger}_{#1}}
\newcommand{\opn}[1]{n^{\phantom{\dagger}}_{#1}} 
\newcommand{\opnq}[1]{\bar{n}^{\phantom{\dagger}}_{#1}}
 
\def\be{\begin{equation}}
\def\ee{\end{equation}}
\def\ba{\begin{eqnarray}}
\def\ea{\end{eqnarray}}

\def\hybrid{\topmargin -10pt    \oddsidemargin 0pt %%%%%%%%%%%%%% Archive-30pt
        \headheight 0pt \headsep 0pt
        \textwidth 16.5cm      % A4 paper
        \textheight 23cm       % A4 paper
        \marginparwidth .875in
        \parskip 5pt plus 1pt   \jot = 1.5ex}

\hybrid

\def\k{\kappa}

\def\a{\alpha}

\def\b{\beta}

\def\d{\delta}
\def\D{\Delta}

\def\m{\mu}

\def\om{\omega}

\def\l{\lambda}
\def\L{\Lambda}
\def\s{\sigma}

\def\no{\noindent}

\def\IR{\relax{\rm I\kern-.18em R}}

\csname @addtoreset\endcsname{equation}{section}

\begin{document}

\begin{titlepage}
\begin{center}
\strut\hfill

{\Large \bf Bethe Ansatz solution of the small polaron with nondiagonal
  boundary terms}

\vskip 0.5in

{\bf Nikos Karaiskos, Andr\'e M. Grabinski and Holger Frahm} \vskip 0.2in

 {Institut f\"ur Theoretische Physik, Leibniz Universit\"at Hannover,\\
Appelstrasse 2, 30167 Hannover, Germany}

\vskip .1in

%\vskip -.15in

{\footnotesize {\tt E-mail: \{nikolaos.karaiskos, andre.grabinski, 
frahm\}@itp.uni-hannover.de}}\\

\end{center}

\vskip 1cm

\centerline{\bf Abstract}

The small polaron with generic, nondiagonal boundary terms is investigated
within the framework of quantum integrability.  The fusion hierarchy of the
transfer matrices and its truncation for particular values of the anisotropy
parameter are both employed, so that the spectral problem is formulated in
terms of a TQ equation.  The solution of this equation for generic boundary
conditions is based on a deformation of the diagonal case.  The eigenvalues of
the model are extracted and the corresponding Bethe Ansatz equations are
presented. Finally, we comment on the eigenvectors of the model and explicitly
compute the eigenstate of the model which evolves into the Fock vacuum when
the off-diagonal boundary terms are switched off.

\no

\vfill

\end{titlepage}
\vfill \eject

\section{Introduction}
The investigation of exactly solvable models over the last many years has been
proven very fruitful in extracting physical information regarding phenomena
lying in strongly coupled regimes \cite{1dimHubbard}.  For integrable models
in one spatial dimension, the framework of quantum inverse scattering method
(QISM) \cite{Faddeev:1979gh, Korepin} provides an efficient way of computing
the various physical quantities. In the present article we study the small
polaron model, which can be also regarded as a graded version of the XXZ
quantum spin chain. The $R$-matrix satisfies the graded Yang-Baxter equation
\cite{Yang_Baxter_eq, Kulish_Sklyanin, Kulish:1985bj} 
\begin{equation}
\label{eqYBE}
R_{12}(\l) \,
R_{13}(\l+\m) \, R_{23}(\m) = R_{23}(\m) \, R_{13}(\l+\m) \, R_{12}(\l) \, ,
\end{equation}
and the resulting bulk Hamiltonian coincides with the bulk Hamiltonian of the
XXZ spin chain, after employing a Jordan-Wigner transformation. The effects of
supersymmetry become visible when periodicity of the boundary conditions is
relaxed, hence more general boundaries are considered, corresponding to open
boundaries in the spin chain picture.

Within a graded version of Sklyanin's reflection algebra
\cite{Sklyanin:1988yz} these boundary conditions can be implemented while
keeping the integrability of the model.  For purely diagonal boundary fields,
i.e.\ respecting the $U(1)$ symmetry of the bulk, the spectrum of the model
can be obtained using algebraic Bethe Ansatz (ABA) methods.
For non diagonal boundary conditions, however, this symmetry is broken and a
simple reference state does not exist anymore, rendering the ABA framework
insufficient.

For the respective ungraded model various methods have been successfully
employed in the past, such as the fusion hierarchy of the transfer matrices
and its truncation \cite{Nepomechie:2001uz, Nepomechie:2003vv}, the
construction of a vacuum state by using gauge transformations \cite{Cao_2002},
and generalized TQ equations \cite{Murgan:2006cx}.  All of these approaches,
however, rely on the boundary parameters satisfying certain algebraic
relations or are restricted to particular values of the bulk anisotropy
parameter. No such constraints on the system parameters appear in solutions
using the representation theory of the $q$-Onsager algebra \cite{BaKo07}, in
terms of functional relations derived directly from the Yang-Baxter algebra
\cite{Galleas:2008zz}, or by separation of variables
\cite{FrSW08,FGSW11,Nicc12}.  So far, an actual solution of the eigenvalue
problem in these formulations is possible for relatively small systems only
and it is unclear how the thermodynamic limit can be approached.

Motivated by a recent analysis of the free fermion case
\cite{Grab_Frahm_free_fermi} where the Grassmannian nature of the nondiagonal
boundary parameters was proven sufficient to solve the model without imposing
any constraints, we studied the small polaron model with nondiagonal boundary
conditions, expecting that a similar situation may also hold. As it will
become transparent in the next sections, it turns out that this is the case
for the interacting fermions as well: supersymmetry lifts the need of imposing
constraints on the boundary parameters.  Furthermore, the structure of the
eigenvectors is greatly restricted, so that certain eigenstates can be
computed exactly for an arbitrary number of chain sites.

The paper is organized as follows: first, we describe the basic facts of the
small polaron model and setup our conventions.
In Section 3 we first recall the functional equations relating the commuting
transfer matrices obtained within the fusion approach obtained in
Ref.~\cite{Grab_Frahm}.  Based on these findings we proceed to derive higher
order functional equations for the transfer matrix of the small polaron model
for particular values of the anisotropy parameter which can be formulated as a vanishing
determinant condition.  The latter allows to formulate the spectral problem in
terms of a TQ equation.  Based on the comparison with a particular limit of
the fusion hierarchy we conjecture that this equation actually describes the
spectrum of the model without any restrictions on the system parameters, i.e.\
anisotropy and boundary fields.  This is supported by the fact that the
solution for diagonal boundary conditions can be shown to be equivalent to
what has been found previously using algebraic or coordinate Bethe ansatz
methods.
In Section 4 we proceed with solving the full nondiagonal model, by deforming
the corresponding problem of the diagonal case. The transfer matrix eigenvalues 
are found to depend on two distinct sets of
Bethe roots, which satisfy two coupled sets of Bethe Ansatz Equations (BAE).
In Section 5 we compute the 'vacuum eigenstate' of the model, i.e.\ the unique
state which reduces to the Fock vacuum in the limit of diagonal boundary
fields.  Based on this result we propose an expression for the generic
structure of the eigenvectors of the model which allows for a complete
solution of the spectral problem in principle.  We conclude with discussing
our results and future directions.

\section{The small polaron with open boundary conditions}
The small polaron model \cite{Fedyanin_1, Fedyanin_2} describes the motion of
an additional electron in a polar crystal.  The physics of these interactions
is captured by the following bulk Hamiltonian
\be
H_\textrm{bulk} = \sum_{j=1}^{N-1} - t\, \big(c^\dagger_{j+1}\, c_j + c^\dagger_j\,  c_{j+1}\big) 
+V \, \big(n_{j+1} \, n_j + \bar{n}_{j+1}\, \bar{n}_j\big) \, ,
\label{Hbulk}
\ee
where $c^\dagger_k$ and $c_k$ denote the creation and annihilation operators
of spinless fermions at site $k$ respectively, obeying anticommutation
relations $\{c_k^\dagger, c_l\} = \d_{kl}$. We have also introduced the
occupation number operators $n_k = c_k^\dagger c_k$ and $\bar{n}_k = 1-n_k$,
so that the parameters $t$ and $V$ may be interpreted as hopping amplitude and
density-density interaction strength respectively. It is possible to derive
the above Hamiltonian through the QISM framework, thus rendering the system
integrable \cite{Pu_Zhao}. In the Hamiltonian (\ref{Hbulk}), periodic boundary
conditions are to be assumed. However, periodicity can be relaxed and one may
consider integrable open boundary conditions, by using the framework of
quantum integrability for models with open boundaries
\cite{Sklyanin:1988yz}. The full Hamiltonian of the system contains the
additional boundary terms, which in general may be non-diagonal ones
\be
H = H_\textrm{bulk} + H_\textrm{diag}+ H_\textrm{nondiag} \, .
\label{Hfull}
\ee
Since we deal with a fermionic lattice model, the local space of states is
$\mathbb{Z}_2$-graded \cite{Gohmann_Murakami}.  The tensor product is graded
according to the rule
\be
(A \otimes_s B)^{ac}_{~~bd} = (-1)^{[p(a)+p(b)]p(c)} A^a_{~b}\,  B^c_{~d} \, ,
\ee
where the parity $p(a)$ is equal to zero (one) for bosonic (fermionic)
indices. All matrix operations below, such as the super trace of a matrix, are
then to be understood as operations on super matrices. We omit the definitions
of these matrix operations and refer the interested reader to the references
\cite{Cornwell, Gohmann_Murakami}, whose conventions we follow.

The fundamental super transfer matrix of the model is given by the super trace
\be
 t(u) = \textrm{str}_0 \Big\{ K^+(u)\, T(u)\, K^-(u)\, \hat{T}(u) \Big\} \, ,
 \label{super_transfer_matrix}
\ee
where the monodromy matrices are defined as
\be
T(u) = R_{N0}(u) \cdots R_{20}(u) \, R_{10}(u) \, , 
\qquad \hat{T}(u) = R_{01}(u)R_{02}(u) \cdots R_{0N}(u) \, .
\label{monodromies_defs}
\ee
The $R$-matrix is given by
\begin{equation}
 R_{ij}(u)=\frac{1}{\sin(2\eta)}
      \begin{pmatrix}
        \sin(u+2\eta)&0&0&0\\
        0&\sin(u)&\sin(2\eta)&0\\
        0&\sin(2\eta)&\sin(u)&0\\
        0&0&0&-\sin(u+2\eta)
      \end{pmatrix}\,
 \label{eq:RMatrix}
\end{equation}
acting on the tensor product $V_i\otimes V_j$ of two linear spaces $\sim
\mathbb{C}^2$.  It satisfies the graded Yang-Baxter equation (\ref{eqYBE}) and
enjoys several useful properties, such as unitarity
\be
R_{12}(u)R_{21}(-u) = \zeta(u) \, ,
\ee
crossing symmetry 
\be
R_{21}^{st_2}(-u-4\eta) R_{21}^{st_1}(u)=\zeta(u+2\eta) \, ,
\ee
and periodicity
\be
R_{12}(u+\pi) = -\s_1^z R_{12}(u) \s_1^z = -\s_2^z R_{12}(u) \s_2^z \, .
\ee
In the above we have also defined
\be 
\zeta(u) \equiv g(u)g(-u), \qquad \textrm{and} \qquad g(u) \equiv 
\frac{\sin(u-2\eta)}{\sin(2\eta)}\, .
\ee

The $K$-matrices, which contain the boundary information, satisfy the graded
reflection algebra \cite{Sklyanin:1988yz, Bracken} and have the following
generic expressions (see also \cite{Zhou, Zhou2, GuanGrimmRoemer})
\begin{equation}
 \begin{aligned}
   K^\sss{-}(u)&=\omega^\sss{-}
          \begin{pmatrix}
            \sin(u+\psi_\sss{-})&\alpha^\sss{-}\sin(2u)\\
            \beta^\sss{-}\sin(2u)&-\sin(u-\psi_\sss{-})
          \end{pmatrix} \,,
%   \label{eq:KMinus}
\\[1em]
   K^\sss{+}(u)&=\omega^\sss{+}
          \begin{pmatrix}
            \sin(u+2\eta+\psi_\sss{+} ) & \alpha^\sss{+}\sin(2u+4\eta)\\
            \beta^\sss{+}\sin(2u+4\eta) & \sin(u+2\eta-\psi_\sss{+})
          \end{pmatrix} \, ,
%   \label{eq:KPlus}
 \end{aligned}
 \label{eq:generalKs}
\end{equation}
with normalizations $\omega^\sss{\pm}$ defined by
\begin{equation}
  \omega^\sss{-} \equiv \frac{1}{\sin\psi_\sss{-}} \qquad \text{and} \qquad
  \omega^\sss{+} \equiv \frac{1}{2\cos2\eta \, \sin\psi_\sss{+}} \, .
\end{equation}
The boundary parameters $\psi_\sss{\pm}$ are commuting numbers with a
non-vanishing complex part, while the parameters $\a^\sss{\pm}, \b^\sss{\pm}$
are odd Grassmann numbers and anticommute:
\be
[\psi_\sss{+}, \, \psi_\sss{-}] = 0 = \{\a^\sss{\pm}, \, \a^\sss{\pm} \} 
= \{\a^\sss{\pm}, \,\b^\sss{\pm} \} = \{\b^\sss{\pm}, \, \b^\sss{\pm} \} \, . 
\ee
Moreover, the reflection algebra is only satisfied provided that the odd
Grassmann numbers are subject to the condition $\a^\sss{+} \cdot \b^\sss{+} =0
= \a^\sss{-}\cdot \b^\sss{-}$.

Through the framework of QISM, the super transfer matrix
(\ref{super_transfer_matrix}) gives rise to a commutative family of conserved
charges, among them the Hamiltonian of the small polaron with open
boundaries, which can be derived as
\be
\left.\frac{d}{du}t(u) \right|_{u=0} = 2H +\textrm{const}\, .
\ee
The bulk part of this Hamiltonian coincides with the expression (\ref{Hbulk})
after identifying $t=-\csc 2\eta$ and $V=\cot2\eta$.  The boundary terms
obtained with (\ref{eq:generalKs}) read as
\be
\begin{split}
& H_{\textrm{diag}} = \mathcal{N}_+ \, \bar{n}_N - \mathcal{N}_-\, n_N +
\frac{1}{2}\cot\psi_\sss{-} \, (\bar{n}_1 -n_1) \cr 
& H_{\textrm{nondiag}} =\csc \psi_\sss{-} \,  ( \a^\sss{-} c_1 - \b^\sss{-}
c^\dagger_1)  
+ \csc\psi_\sss{+} \, ( \a^\sss{+} c_N - \b^\sss{+} c^\dagger_N) \, ,
\end{split}
\label{H_boundary}
\ee
where the following shorthands have been also introduced:
$\mathcal{N}_\pm\equiv\frac{1}{2}\csc(2\eta) \csc(\psi_\sss{+})
\sin(2\eta\pm\psi_\sss{+})$.
The diagonal boundary terms can be identified with static boundary chemical
potential at the first and last site of the lattice, respectively.  By means
of a Jordan-Wigner transformation the bulk and diagonal boundary terms can be
mapped to spin-1/2 XXZ Heisenberg chain with boundary magnetic fields, see
also Appendix A.  This is not possible for the off-diagonal terms breaking the
$U(1)$ symmetry of the system.  In the small polaron formulation the terms in
$H_{\textrm{nondiag}}$ can be interpreted as sources and sinks for injection
of additional particles into the system.  Their amplitudes are odd Grassman
numbers, reflecting the fermionic nature of the corresponding reservoir.  The
Jordan-Wigner transformation of these terms yields a non-local expression in
the spin chain formulation.

The super transfer matrix enjoys crossing symmetry
\be 
t(-u-2\eta) = t(u) \, ,
\label{trns_crs_symm}
\ee
and periodicity
\be
t(u+\pi) = t(u)\, . 
\ee
For future use, we also note that the open transfer matrix is normalized as
$t(0) = \mathds{1} $ and becomes diagonal in the semi-classical limit
$\eta\to0$
\ba
&& t(u)\Big|_{\eta=0} = \frac{(\sin u)^{2N}}{\sin\psi_\sss{+}\sin\psi_\sss{-}} \Big[ 2\sin^2u  \cos^2u \, (\b^\sss{+}\a^\sss{-} - 
\a^\sss{+}\b^\sss{-}) \, \prod_{k=1}^N \s^z_k \cr 
&& \hspace{3cm} - \big(\cos^2u \, \sin\psi_\sss{-}\sin \psi_\sss{+}  +\sin^2u \cos\psi_\sss{-} \cos\psi_\sss{+}\big) \,  
\mathds{1} \Big]\, .
\label{transfer_semiclassical_limit}
\ea
Finally, taking the limit $z\equiv e^{iu} \to \infty$, the asymptotic behavior
of the super transfer matrix is obtained, which is needed for later
comparison. The leading term contains only odd Grassmann boundary parameters
and has the expression
\be
t(z)  =  \left(\frac{z}{2i\sin 2\eta}\right)^{2N} \frac{\om^\sss{+} \om^\sss{-}}{4}\,   z^4 \, e^{4i\eta} \, 
 (\b^\sss{+}\a^\sss{-} - \a^\sss{+}\b^\sss{-})  \, \prod_{j=1}^N(\bar{n}_j - e^{2i\eta}n_j)(n_j+e^{2i\eta}
 \bar{n}_j) + \mathcal{O}(z^2)\, .
 \label{asymptotic_transfer}
\ee
This particular combination of the odd Grassmann boundary parameters emerges
in many different relations, and as will become transparent below it appears
uniquely in all eigenvalues of the transfer matrix, thus henceforth it will be
denoted as
\be
\mathcal{G} \equiv \b^\sss{+}\a^\sss{-} - \a^\sss{+}\b^\sss{-} \, .
\label{mathcalg}
\ee

\section{Functional relations}
As stated in the introduction our analysis of the spectral problem for the
transfer matrix of the small polaron model relies on its reformulation in
terms of functional equations.  To keep the presentation of our results
self-contained we begin by recalling the results of Ref.~\cite{Grab_Frahm},
namely the infinite hierarchy of fused transfer matrices and its truncation
for particular values of the anisotropy parameter $\eta$.  Then we use the
truncation identity to rewrite the fusion hierarchy as a higher order
functional relation for the transfer matrix (\ref{super_transfer_matrix})
alone.  This relation is then shown to be equivalent to a vanishing
determinant condition which allows for the formulation of the spectral problem
in terms of a TQ equation.  We then argue that this TQ equation holds for
arbitrary values of the anisotropy and generic non-diagonal boundary
conditions.

\subsection{Fusion hierarchy and truncation identities}
Having a transfer matrix at hand, one may construct a family of commuting
transfer matrices derived from auxiliary spaces of higher dimension, through a
fusion procedure \cite{Kulish:1981gi, Mezincescu:1991ke}.  Choosing a suitable
normalization the fused transfer matrices form a hierarchy related by the
recursion relations 
\begin{equation}
  t^{(n)}(u) \cdot t^{(1)}(u+n\cdot 2\eta) =
  t^{(n+1)}(u) -  
          \tilde{\Delta}(u+[n-1]\cdot 2\eta) \cdot t^{(n-1)}(u) \, .
  \label{eq:OBCFusionHierarchyNoTilde}
\end{equation}
with $t^{(0)} \equiv \mathds{1}$ and $t^{(1)} \equiv -t(u)$.
The function $\tilde\Delta(u)$ is up to a scaling factor  the
super quantum determinant of the small polaron model \cite{Grab_Frahm}
\begin{equation}
\begin{aligned} 
  \tilde\D(u)\,\zeta(2u+4\eta) &\equiv \D(u)\\ & =
  \zeta^{2N}(u+2\eta) \, g(-2u-6\eta) \, g(2u+2\eta) \det
  K^+(u) \det K^-(u+2\eta) \, . 
\end{aligned}
\end{equation}
Note that the quantum determinant does not depend on the off-diagonal elements
of the boundary matrices $K^\pm(u)$ as a consequence of the nilpotency of the
odd Grassmann parameters.
Since the fused transfer matrices commute with each other, the fusion
hierarchy can be also read as a set of relations between their eigenvalues,
which after shifting the spectral parameter $u\to u-n\cdot 2\eta$ gives
\begin{equation}
\label{FUS-n}
 -\L(u) = \frac{\L^{(n+1)}(u-n\cdot
  2\eta)}{\L^{(n)}(u-n\cdot 2\eta)} - \tilde{\D}(u-2\eta) 
\frac{\L^{(n-1)}(u-n\cdot 2\eta)}{\L^{(n)}(u-n\cdot 2\eta)}\, .
\end{equation}
For later use we rewrite these equations in terms of the functions
\begin{equation}
  \L^{(n)}(u-n\cdot2\eta) 
  \equiv Q^{(n)}(u) \prod_{k=0}^n \kappa(u-k\cdot 2\eta) \, ,
\end{equation}
giving
\begin{equation}
\label{TQeq-n}
\begin{aligned}
  -\L(u) &= h^+(u) \frac{Q^{(n+1)}(u+2\eta)}{Q^{(n)}(u)}
                      - h^-(u) \frac{Q^{(n-1)}(u-2\eta)}{Q^{(n)}(u)}\,,\\
  & h^+(u) = \kappa(u+2\eta)\,,\qquad 
    h^-(u) = \tilde{\D}(u-2\eta)\frac{1}{\kappa(u)}\,.
\end{aligned}
\end{equation}
Note that while the coefficient functions $h^{\pm}(u)$ can be modified by
choosing different factors $\kappa(u)$ in the definition of $Q^{(n)}(u)$ they
always factorize the rescaled quantum determinant, i.e.\ $h^+(u) h^-(u+2\eta)
= \tilde{\D}(u)$.

The functional equations (\ref{FUS-n}) constitute a set of relations for an
infinite set of unknown functions $\tilde{\Lambda}^{(n)}(u)$.  Restricting the
anisotropy parameter to 'roots of unity', $\eta_n=\frac{\pi/2}{n+1}$, is has
been shown that the fused transfer matrices at level $n+1$ and $n-1$ are related
by the truncation identity \cite{Grab_Frahm}
\begin{equation}
t^{(n+1)}(u,\eta_{n}) =  \phi^{\text{id}}_{n}(u) \cdot \mathbb{I} -
\phi^{\tau}_{n}(u)  
\cdot t^{(n-1)}(u+2\eta_{n},\eta_{n}) \, ,
\label{eq:obcTauTruncation}
\end{equation}
where the functions $\phi^{\text{id}}_{n}(u), \phi^{\tau}_{n}(u)$ are given by
the following expressions 
\ba 
&& \phi^{\text{id}}_{n}(u)  =  \mathcal{M}_{n}^{2N}(u)\ \mu^{\sss
  +}_{n}(u)\mu^{\sss -}_{n}(u) 
[ \nu^{\sss +}_{n}(-u)\nu^{\sss -}_{n}(u) + \nu^{\sss +}_{n}(u)\nu^{\sss
  -}_{n}(-u) ] \nonumber \\ 
&& \phi^{\tau}_{n}(u) = \zeta^{2N}(u) \m_n^+(u) \m_n^-(u) \, ,
\ea
with
\begin{eqnarray}
\mathcal{M}_n(u) & \equiv & \left(\frac{1/2}{\sin2\eta_n}\right)^n 
\frac{\sin([n+1]u)}{\sin2\eta_n} \cr
\mu^\sss{\pm}_n(u) &\equiv& \pm \delta\{K^\sss{\pm}(\mp
u-2\eta_n,\eta_n)\}\frac{\sin(2\eta_n)}{\sin(2u-2\cdot 2\eta_n)} 
\prod_{k=2}^{2n}\frac{\sin(2u+k\cdot 2\eta_n)}{\sin(2\eta_n)}\cr
\nu^\sss{\pm}_n(u) &\equiv& \mp\frac{\omega_n^\sss{\pm}}{\mu^\sss{\pm}_n(u)}
\left(\frac{\omega_n^\sss{\pm}}{2}\right)^n 
\sin([n+1][u\mp\psi_\sss{\pm}])
\prod_{i=1}^n \prod_{j=1}^i \frac{\sin(2u+[i+j]\cdot 2\eta_n)}{\sin(2\eta_n)}\ .
\end{eqnarray}

\subsection{Higher order functional equation and determinant representation}
Combining the fusion hierarchy with the truncation identity for a particular
value $n$, yields an $(n+1)$-order functional equation for the transfer matrix
of the model with the corresponding anisotropy $\eta=\eta_n$. Unfortunately,
it does not seem possible to write the functional relations in a closed form.
Starting from low values of $n$, one observes that at each level new terms
emerge and their number is given by the Fibonacci numbers $F(n)$. In
particular, at a given value $n$, there are
\[
F(n+1) + F(n-1) +1 ~, \qquad n=2,3, \cdots \, 
\]
terms in total. Nevertheless, after carefully examining the structure of the
functional relations, it is seen that the following simple schematic structure
appears
\begin{equation}
 \label{funct_rel}
 \begin{aligned}
   & t(u) \,  t(u+2\eta) \cdots t(u+(2n-2)\eta) + \mho_1
   + \phi^\tau_{n-1}(u) \cdot \mho_0 \\
   & \qquad = \phi_{n-1}^{id}(u) + \prod_{k=0}^{\frac{n}{2}-1}
   \tilde{\Delta}(u+4k\eta) 
   + \, \phi^\tau_{n-1}(u) \, \prod_{k=1}^{\frac{n}{2}-1}
   \tilde{\Delta}(u+(4k-2)\eta)\,. 
 \end{aligned}
\end{equation}
First, we should note that the last two terms in the RHS are present only for
even values of $n$. Next, the symbol $\mho$ stands for the following sequence:
\ba
 && \mho_1 :=   \sum_{q_0 = 0}^{n-2} \, \tilde{\Delta}(u+2q_0 \eta) ~ \mathcal{T}_{q_0}(u) \nonumber \\ 
 && \qquad +  \sum_{q_1=0}^{n-4} \, \tilde{\Delta}(u+ (2q_1+4) \eta) 
 \sum_{q_0=0}^{q_1} \, \tilde{\Delta}(u+2q_0 \eta) ~ \mathcal{T}_{q_0q_1}(u) \nonumber \\ 
 && \qquad +  \sum_{q_2=0}^{n-6}\, \tilde{\Delta}(u+ (2q_2+8) \eta)  \sum_{q_1=0}^{q_2} \, 
 \tilde{\Delta}(u+ (2q_1+4) \eta)
 \sum_{q_0=0}^{q_1} \tilde{\Delta}(u+2q_0 \eta) ~ \mathcal{T}_{q_0q_1q_2}(u) \nonumber \\ 
 & & \qquad + \cdots \, ,
\label{om_seq}
\ea
with 
\[
 \mathcal{T}_{q_0q_1\cdots q_{\ell}}(u) \equiv \prod_{\substack{m=0 \\ m\notin
     \mathcal{Q}}}^{n-1} t (u+2m\eta),  
 \qquad \mathcal{Q}= \bigcup_{k=1}^\ell  \{q_k +2k, q_k+ 2k +1\}\,.
\]
A similar structure holds for $\mho_0$ as well, although some lower/upper
limits of the sums and the product are different. Nevertheless, the structure
above is reminiscent of a path-ordered exponential, and appears as some
deformed discrete version of the latter one. It would be interesting to see if
this observation possesses some physical meaning, i.e., if the corresponding
operator in the path-ordered exponential plays some physical role here.

Despite the fact that the functional relations cannot be written in a closed
form, they can be cast into a vanishing determinant representation as a
consequence of the
restriction $m \notin \mathcal{Q}$ in Eq.~(\ref{om_seq}) above.
This
leads to the derivation of a TQ equation for the eigenvalues of the transfer
matrix (\ref{super_transfer_matrix}) \cite{Bazhanov:1987zu}. For the small
polaron model, it turns out that the vanishing determinant representation has
the same structure with the corresponding one of the XXZ model
\cite{Nepomechie:2003vv}. In particular, the functional relations
(\ref{funct_rel}) can be also written in the following form
\be
\det 
\begin{pmatrix}
  \L_0 & -\tilde{h}_{-1} & 0 & \cdots &  0& -h_0 \cr
 -h_1 & \L_1 & -\tilde{h}_0 & 0  & \cdots   & 0 \cr
  0 & -h_2 & \L_2 & -\tilde{h}_1 & \ddots & \vdots \cr
  \vdots & 0 & \ddots  & \ddots & \ddots & 0\cr 
  0 & \vdots  & \ddots & -h_{n-1} & \L_{n-1} & -\tilde{h}_{n-2} \cr
  -\tilde{h}_{n-1}  & 0 & \cdots & 0 & -h_n & \L_n
\end{pmatrix} = 0~,
\label{det_rep}
\ee
where $\L_k \equiv \L(u+2k\eta)$ is an eigenvalue of
$t(u+2k\eta)$, $~h_{k} \equiv h(u+2k\eta)$ and $\tilde{h}_k \equiv
\tilde{h}(u+2k\eta)$.  At this point, the functions $h_k, \tilde{h}_k$ are
unknown which have to be determined by requiring relations (\ref{funct_rel})
and (\ref{det_rep}) to be identical.  This gives
\begin{equation}
   \begin{split}
    h(u+2\eta) \, \tilde{h}(u-2\eta) & =   -\tilde{\Delta}(u)\,, \\
    h(u) \, \tilde{h}(u+ 2(n-1) \eta) & =   -\phi^\tau_n(u)\,, \\
    \prod_{k=0}^n h(u+2k\eta) + \prod_{k=-1}^{n-1} \tilde{h}(u+2k\eta) & =
    \phi^{id}_{n}(u)\,. 
  \end{split}
  \label{condtosolve}
\end{equation}
Inspired by the structure of the $h$-functions in the XXZ case
\cite{Nepomechie:2003vv}, as well as the functions found in \cite{Grab_Frahm},
we consider the following expressions
\be
 \tilde{h}(u) = h(-u-4\eta), \qquad h(u) = \left(\frac{\sin(u+2\eta)}{\sin 2\eta}\right)^{2N} 
 \frac{\sin(2u+4\eta)}{\sin(2u+2\eta)} g_-(u)g_+(u) \, ,
 \label{h_trnc_interm}
\ee
where the boundary information is contained in the functions $g_\pm(u)$. Substituting
the above expressions into (\ref{condtosolve}), the following condition arises
regarding the functions $g_{\pm}(u)$:
\be
 g_-(u +2\eta )g_-(-u-2\eta )g_+(u+2\eta)g_+(- u - 2\eta) = - \det K^+(u) \det K^-(u+2\eta) \, .
\ee
Assuming that the functions $g_{\pm}(u)$ factorize the determinants of the 
reflection matrices and recalling the explicit expressions for the determinants of the 
boundary matrices, one obtains
\begin{equation}
  \begin{aligned}
    & g_-(u )g_-(-u) = \det K^-(u) = -(\om^\sss{-})^2 \sin(u  +
    \psi_\sss{-})\sin(u - \psi_\sss{-}) \\ 
    & g_+(u)g_+(- u ) = - \det K^+(u-2\eta) = -(\om^\sss{+})^2 \sin(u +
    \psi_\sss{+})\sin(u - \psi_\sss{+}) \, , 
  \end{aligned}
\end{equation}
pinpointing to the natural solutions
\be
g_-(u) = \om^\sss{-} \sin(u+\psi_\sss{-})\, , \qquad 
g_+(u) = \om^+ \sin(u+\psi_\sss{+}) \, .
\label{gsol}
\ee
Using the expressions (\ref{h_trnc_interm}) and (\ref{gsol}) then, it is
straightforward to check that all three relations in (\ref{condtosolve}) are
automatically satisfied, so that the expressions for $h(u), \tilde{h}(u)$ are
now proven. It is interesting to point out that the structure we have found
here is identical with the corresponding structure of the purely diagonal XXZ
case \cite{Nepomechie:2003vv}.

The vanishing determinant guarantees the existence of a non-trivial null
vector, $(Q_0, Q_1, \cdots Q_n)$, so that the following relations hold
\ba
\L_0 Q_0 - \tilde{h}_{-1} Q_1 - h_0 Q_n & = & 0 \, , \nonumber \\
-h_k Q_{k-1} + \L_k Q_k - \tilde{h}_{k-1} Q_{k+1} & = & 0 \, , \qquad
k=1, \cdots, n-1 \, ,\\ 
-\tilde{h}_{n-1} Q_0 - h_{n} Q_{n-1} + \L_n Q_n & = & 0 \, . \nonumber
\ea
For the present choice of the quasi-classical parameter, i.e.\
$\eta=\eta_n=\frac{\pi/2}{n+1}$, these equations can be recast as TQ equations
for a periodic function $Q(u)=Q(u+\pi)$ by identifying $Q_k \equiv Q(u+2\eta
k)$:
\be
 \L(u) = h(u) \frac{Q(u-2\eta)}{Q(u)} + h(-u-2\eta)
 \frac{Q(u+2\eta)}{Q(u)}\,, 
 \quad \mathrm{for~}\eta=\eta_n. 
\label{fnctrel_tq_eq}
\ee
Here the degree of the Fourier polynomial for the function $Q(u)$ depends on
the choice of the anisotropy $\eta=\eta_n$ (or, equivalently, the level of the
fusion hierarchy where the truncation identity (\ref{eq:obcTauTruncation})
appears).  We note that as a consequence of Eqs.\ (\ref{trns_crs_symm}) and
(\ref{fnctrel_tq_eq}) the function $Q(u)$ is crossing symmetric.

We emphasize that up to this point we have made no assumptions concerning the
form of the reflection matrices $K^\pm(u)$.  In particular the functional
equation (\ref{fnctrel_tq_eq}) holds for both diagonal and non-diagonal
boundary conditions and coincide with the limit $n\to\infty$ of
Eq.~(\ref{TQeq-n}) of the fusion hierarchy provided that the limit
$\lim_{n\to\infty} Q^{(n)}(u) = Q(u)$ exists \cite{Yang:2005ce}.  The first of the conditions
(\ref{condtosolve}) used to derive the determinant representation of the
functional equation guarantees the factorization $h^+(u) h^-(u+2\eta) =
-h(u+2\eta) h(-u-2\eta) = \tilde\Delta(u)$ of the quantum determinant with
\begin{equation}
  h^-(u) = -h^+(-u-2\eta) = -\omega^+\omega^- \left(\frac{\sin(u+2\eta)}{\sin2\eta}\right)^{2N}
  \frac{\sin(2u+4\eta)}{\sin(2u+2\eta)} \sin(u+\psi_+)\sin(u+\psi_-) \, ,
  \label{hfunctions}
\end{equation}
corresponding to 
\begin{equation}
  \label{kappaTQ}
  \kappa(u) = \omega^- \omega^+ \left(-\frac{\sin(u-2\eta)}{\sin2\eta}\right)^{2N} \frac{\sin(2u-4\eta)}{\sin(2u-2\eta)}
  \sin(u-\psi_-)\sin(u-\psi_+) \, ,
\end{equation}
in the definition of the $Q^{(n)}(u)$.  

For the limit $\lim_{n\to\infty} Q^{(n)}(u)$ to exist, however, this choice of
the function $h(u)$ is not sufficient for general boundary conditions: in the
following Section we shall introduce deformations of the functions $h$ and $Q$
which allow for an algebraic solution of the TQ equation (\ref{fnctrel_tq_eq})
in the case of non-diagonal boundary conditions.

In the case of particle number conserving \emph{diagonal} boundary matrices
the TQ equation can be solved in terms of Fourier polynomials of fixed degree
for the $Q$-functions.  Comparing to what is obtained using the algebraic or
coordinate Bethe ansatz they are given by \cite{Umeno_Fan_Wadati, Fan_Guan, GuanFanYang}
\be
  Q(u) = \prod_{\ell=1}^M \sin(u-v_\ell^{(0)}) \sin(u+v_\ell^{(0)} +2\eta) \, ,
  \label{Q_fnc_diag}
\ee
for all anisotropies $0\le\eta\le \pi/2$.
For small system sizes $N$ we have checked by explicit computation of the
functions $Q^{(n)}(u)$ from the recursion relations (\ref{TQeq-n}) for
diagonal boundary fields that they do in fact converge to the expression
(\ref{Q_fnc_diag}) for $n\to\infty$ for all anisotropy parameters.
The zeros $v_\ell^{(0)}$ of the $Q$-functions are determined by requiring
analyticity of the eigenvalues $\Lambda(u)$ of the transfer matrix from
the TQ equation (\ref{fnctrel_tq_eq}).  This yields the BAE for the diagonal
model
\be
\frac{h^-(v_\ell^{(0)})}{h^+(v_\ell^{(0)})} = \frac{Q(v_\ell^{(0)} +
  2\eta)}{Q(v_\ell^{(0)} - 2\eta)} \, , 
\ee
or upon substitution 
\ba &&
\left(\frac{\sin(v_\ell^{(0)} + 2\eta)}{\sin v_\ell^{(0)}}\right)^{2N}
\frac{\sin(v_\ell^{(0)}+\psi_\sss{-})\sin(v_\ell^{(0)}+\psi_\sss{+})}{\sin(v_\ell^{(0)}+2\eta-\psi_\sss{+})
  \sin(v_\ell^{(0)}+2\eta - \psi_\sss{-})} = \nonumber \\
&&\hspace{4cm} = \prod_{j\neq \ell}^M \frac{\sin(v_\ell^{(0)} - v_j^{(0)}
  +2\eta)\sin(v_\ell^{(0)} + v_j^{(0)} + 4\eta)} {\sin(v_\ell^{(0)} -v_j^{(0)}
  - 2\eta)\sin(v_\ell^{(0)} + v_j^{(0)})} \, .
\label{BAE_diag}
\ea

\section{Non-diagonal boundaries}
Having reporduced the spectrum of the diagonal model, we now turn our
attention to the case where nondiagonal boundary conditions are considered.
Based on our findings above we assume that the spectral problem is given by
the TQ equation (\ref{fnctrel_tq_eq}) for arbitrary values of the anisotropy
$\eta$.
It turns out that the incorporation of the nondiagonal contributions can be
done efficiently by appropriately deforming the $h$ and $Q$-functions of the
diagonal model.  Thus our starting point would be a deformation of the TQ
equations themselves
\begin{equation}
\label{TQeq}
  \L(u) = H^-(u) \frac{\tilde{Q}(u-2\eta)}{\tilde{Q}(u)} 
  - H^+(u) \frac{\tilde{Q}(u+2\eta)}{\tilde{Q}(u)} \, .
\end{equation}
where $\L(u)$ denote the eigenvalues of the full, nondiagonal problem. 
Regarding the $H$-functions first, since the diagonal case should be contained 
within the construction as a
special limit, we propose the following deformations 
\be
H^{\pm}(u) = h^{\pm}(u) \big(1 + \mathcal{G} \, f^{\pm}(u)\big) \, ,
\label{H_deform}
\ee
where $h^{\pm}(u)$ are given in (\ref{hfunctions}), $\mathcal{G}$ is the
combination of the odd Grassmann parameters defined in (\ref{mathcalg}) and
$f^{\pm}(u)$ are generic functions to be determined. There exist two
constraints on the expressions of these unknown functions. First, their
contribution in the factorization of the quantum determinant should vanish,
yielding
\be
f^+(u-2\eta) = - f^-(u) \, .
\ee
Furthermore, the crossing symmetry of the einvalues $\L(u)$ entailed
by (\ref{trns_crs_symm}) should be preserved.  We therefore assume that the 
$\tilde{Q}$-functions enjoy crossing symmetry as well, similarly to the
functional equation (\ref{fnctrel_tq_eq}) derived from the truncation
identities.   This implies that the $H$-functions are related
through $H^-(u) = - H^+(-u-2\eta)$. This
relation holds automatically for the $h$-functions of the diagonal case,
whereas it also provides a second constraint on $f^\pm(u)$:
\be
f^+(-u-2\eta) = f^-(u) \, .
\label{2nd_cnstrnt_f}
\ee
A set of solutions to these functional relations is given by
\ba
f^+(u) & = & \mathcal{W} \, \sin(2u+4\eta) \cr
f^-(u) & = & -\mathcal{W} \, \sin(2u) \, , 
\label{ffunct}
\ea
with the coefficient $\mathcal{W}$ to be determined from the asymptotic behavior (\ref{asymptotic_transfer}) 
of the transfer matrix eigenvalues. It should be stressed
out that the above choice is not unique, but it was found to be the only 
one consistent with the various limits and with the constraints arising from the
functional relations. 

A similar deformation should be considered for the $\tilde{Q}$-functions as
well, since they are assumed to originate from the higher spin eigenvalues
through a limiting procedure. The most general ansatz extending the functions
$Q(u)$ of the diagonal case (\ref{Q_fnc_diag}) and containing the special
combination of Grassmann numbers, $\mathcal{G}$, reads as \be \tilde{Q}(u) =
Q(u) + \mathcal{G} \, b(u) \, ,
\label{Q_deform}
\ee 
with $b(u)$ being a function to be determined. Substituting the deformations
(\ref{H_deform}) and (\ref{Q_deform}) into the TQ equation (\ref{TQeq}), the latter 
one becomes
\ba
&& \L(u) =  \L^{\textrm{diag}}(u)\left(1 - \mathcal{G} \, \frac{b(u)}{Q(u)}\right)  
 + \mathcal{G}\left[h^-(u)\left( \frac{b(u - 2\eta)}{Q(u)} + f^-(u)\frac{Q(u-2\eta)}{Q(u)}\right) \right.  \nonumber \\
&& \hspace{3cm} \left. -h^+(u)\left( \frac{b(u+2\eta)}{Q(u)} + f^+(u)\frac{Q(u+2\eta)}{Q(u)} \right)\right] \, .
\label{TQnd}
\ea
Additional requirements are needed to
determine the function $b(u)$ in (\ref{Q_deform}).  Again the choice of this
function has to guarantee the analyticity of the additional terms appearing in
the functional equation (\ref{TQnd}) for the eigenvalues. In addition, $b(u)$
should enjoy crossing symmetry, which was assumed in order to derive the 
constraint (\ref{2nd_cnstrnt_f}).
In this spirit, we propose that the nondiagonal correction to the
$\tilde{Q}$-functions is given by
\be
b(u) = \prod_{\ell=1}^{M'} \sin(u-v_\ell^{(1)})\sin(u+v_\ell^{(1)}+2\eta) \, ,
\label{b_funct}
\ee
with $\{v_\ell^{(1)} \} \neq \{v_\ell^{(0)}\}$ in general.  Matching of the
asymptotic behavior dictates that the upper limits of the products in
eqs. (\ref{Q_deform}) and (\ref{b_funct}) should be equal, $M'=M$. Furthermore, the 
asymptotics (\ref{asymptotic_transfer}) provide the explicit expression for the
coefficient $\mathcal{W}$ as
\be
 \mathcal{W} = \frac{1}{\sin[\psi_\sss{+} + \psi_\sss{-} + (N-2M-1)2\eta]} \, .
 \label{wcoef}
\ee
Apart from the asymptotics, one should also consider additional limits of the
TQ equation in order to check the consistency of the procedure. The limit
$u\to0$ gives $\L(0) = 1$, which stems from the normalization of the
transfer matrix. Setting $u=0$ into the TQ equation, after a quick inspection
one is lead to the following constraint for $b(u)$
\be
 b(-2\eta) = b(0)\, ,
\ee
which is a special case of a crossing symmetry requirement and automatically
satisfied by the choice (\ref{b_funct}).  Finally, it is interesting to
consider the semi-classical limit $\eta\to 0$. In this limit the $b(u)$
functions drop out from the TQ equation and the eigenvalues become
\be
\L(u)\Big|_{\eta\to0} = \big(h^-(u) - h^+(u)\big) + \mathcal{G}\,  f^-(u) \big(h^-(u) + h^+(u)\big) \, ,
\ee
finding perfect agreement with the semi-classical limit obtained from the
transfer matrix (\ref{transfer_semiclassical_limit}), after recalling the
expressions for $h^\pm(u)$ (\ref{hfunctions}) and $f^\pm(u)$ (\ref{ffunct}) 
with the coefficient $\mathcal{W}$ (\ref{wcoef}) for $\eta\to0$.

The parameters $v_\ell^{(1)}$ in (\ref{b_funct}) are determined by requiring
analyticity of the eigenvalues of the full transfer matrix.  The purely complex part of 
the TQ equation (\ref{TQnd}) yields the
already derived BAE (\ref{BAE_diag}), while the one containing Grassmann
numbers provides a second set of relations, involving both sets of roots
$v^{(0)}$, $v^{(1)}$, which reads as
\be
 \frac{h^-(v_\ell^{(0)})}{h^+(v_\ell^{(0)})} = \frac{\L^{\textrm{diag}}(v_\ell^{(0)}) \, b(v_\ell^{(0)})}
{h^+(v_\ell^{(0)})\left(b(v_\ell^{(0)}-2\eta)+f^-(v_\ell^{(0)})Q(v_\ell^{(0)}-2\eta)\right)}
+ \frac{b(v_\ell^{(0)}+2\eta) + f^+(v_\ell^{(0)}) Q(v_\ell^{(0)}+2\eta)}
{b(v_\ell^{(0)}-2\eta)+f^-(v_\ell^{(0)})Q(v_\ell^{(0)}-2\eta)} \, .
\label{BAE_nondiag}
\ee
Note that the roots $v^{(0)}$ and $v^{(1)}$ also enter this set of equations
through the functions $Q(u)$ and $b(u)$.  

In summary, we have two sets of algebraic equations fixing the parameters
appearing in the ansatz for $\tilde{Q}(u)$.  These equations are of nested
Bethe Ansatz type, similar to those appearing in integrable models based on
higher rank symmetries. It should be stressed out that the 
expression for the eigenvalues of the super transfer matrix (\ref{TQnd}) and the 
corresponding BAE (\ref{BAE_nondiag}) rely on the conjectural deformations of 
the $h$ and $Q$-functions in equations (\ref{H_deform}) and (\ref{Q_deform}) 
respectively. However, we have explicitly verified that our proposed scheme is compatible 
with the various limits and symmetry requirements, namely the truncation to the 
diagonal case, the crossing symmetry, the limits $u\to0$ and $\eta\to 0$ and 
the asymptotics of the transfer matrix. Moreover, the deformations of the $h$-
functions are compatible with the constraints arising from the functional 
relations, namely eqs. (\ref{condtosolve}), which provides an additional, nontrivial
check of our expressions. In order to further enforce the validity of  
our proposed scheme for the eigenvalues as computed from the TQ equation, we 
have  analytically diagonalized the transfer matrix for a small number of chain 
lengths. The comparison of the exact eigenvalues with the ones obtained from 
(\ref{TQnd}) exhibits perfect agreement.

\section{The eigenstates of the model} 
In the previous sections we have obtained the eigenvalues of the transfer
matrix from the the fusion hierarchy and the resulting TQ equation.  Similar
as in the case of diagonal boundary matrices the eigenvalues can be associated
to sectors labelled by the integer $M$ of parameters appearing in the
solution.  This is remarkable since the related $U(1)$ symmetry is broken when
the nondiagonal boundary conditions are applied.

As is common to functional Bethe Ansatz approaches our solution of the
spectral problem thus far does not provide information regarding the
eigenstates of the model.
However, exploiting the existence of the odd Grassmann numbers and their
nilpotency, it is possible to exactly compute the $M=0$ state of the model for
an arbitrary number $N$ of chain sites, solely by using the derived
expressions of the eigenvalues.  To this end we choose $k$-particle
states, $k=0,\ldots,N$,
\be
|i_1i_2\cdots i_k\rangle \equiv c^\dagger_{i_1}c^\dagger_{i_2} \cdots
c^\dagger_{i_k } | \Omega\rangle   \, ,  \qquad
1\le i_1<i_2<\ldots<i_k\le N
\ee
as basis of the Hilbert space of the system, $| \Omega\rangle$ is the Fock
vacuum of the system (and the $M=0$ eigenvector of the diagonal problem).

The key observation used for the construction of the unique $M=0$ eigenvector
of the model with nondiagonal boundary conditions is that for generic chain
length $N$, only the Fock vacuum and basis states containing one or two
particles contribute with a non-vanishing amplitude, all other sectors
decouple.  This allows to express the $M=0$ eigenvector of the nondiagonal
model as 
\be |\Psi \rangle_{M=0} = |\Omega \rangle +
\b^\sss{+} \sum_{i=1}^N b^\sss{+}_i |i\rangle + \b^\sss{-} \sum_{i=1}^N
b^\sss{-}_i |i\rangle +\b^\sss{+}\b^\sss{-} \sum_{i<j}^N B_{ij} |ij \rangle \,
.
\label{ground_state_ansatz}
\ee
The eigenvalue problem for the complete Hamiltonian reads as
\be
H |\Psi\rangle = (\l_\textrm{diag} + \mathcal{G}\, \l_{\textrm{nondiag}})
|\Psi\rangle\, . 
\ee
Splitting the terms with respect to the order of $\b^\pm$ we obtain
\ba
\mathcal{O}(\b): && (H_\textrm{bulk} + H_\textrm{diag}) \, \b^\k \sum_i b^\k_i |i\rangle + 
H_\textrm{nondiag} |\Omega\rangle = \l_\textrm{diag} \, \b^\k \sum_i b^\k_i |i\rangle\cr
\mathcal{O}(\b^2): &&  (H_\textrm{bulk}+H_\textrm{diag}) \, \b^+ \b^- \sum_{i<j} B_{ij} |ij\rangle + 
H_\textrm{nondiag} \, \b^\k \sum_i b^\k_i |i\rangle = \cr
&& \hspace{4cm} = \mathcal{G} \, \l_\textrm{nondiag}|\Omega\rangle + \l_\textrm{diag} \, 
\b^\sss{+} \b^\sss{-} \sum_{i<j} B_{ij}|ij\rangle\, .
\ea
Regarding the terms linear in $\b^\sss{\pm}$ first, after the relevant computations and splitting the resulting
relations with respect to the appropriate excited states, one ends up with the following six 
relations that determine the coefficients $b_i^\sss{\pm}$
\ba
&& \b^\sss{+}|1\rangle: \qquad 
-t\, b^\sss{+}_2 + b_1^\sss{+} \big[V(N-2) - \tfrac{1}{2}\cot\psi_\sss{-} + \mathcal{N}_\sss{+}  -\l_\textrm{diag}\big]  = 0 \cr
&& \b^\sss{+}|N\rangle: \qquad -t\, b^\sss{+}_{N-1} - \csc \psi_\sss{+} 
+ b_N^\sss{+} \big[V(N-2) + \tfrac{1}{2}\cot\psi_\sss{-} -\mathcal{N}_\sss{-} - \l_\textrm{diag}\big]  = 0 \cr
&& \b^\sss{-}|1\rangle: \qquad 
-t\, b^\sss{-}_2 - \csc\psi_\sss{-} + b_1^\sss{-} \big[V(N-2) - \tfrac{1}{2}\cot\psi_\sss{-} 
+ \mathcal{N}_\sss{+} - \l_\textrm{diag}\big]  = 0 \cr
&& \b^\sss{-}|N\rangle: \qquad 
-t\, b^\sss{-}_{N-1} + b_N^\sss{-} \big[V(N-2) + \tfrac{1}{2}\cot\psi_\sss{-} 
-\mathcal{N}_\sss{-} - \l_\textrm{diag}\big]  = 0 \cr
&& \b^\sss{\pm} |\ell\rangle: \qquad -t \, ( b_{\ell-1}^\sss{\pm} +  b_{\ell+1}^\sss{\pm} ) 
+ b_\ell^\sss{\pm} \big[ V(N-3) + \tfrac{1}{2}\cot\psi_\sss{-} 
+\mathcal{N}_\sss{+} - \l_\textrm{diag} \big] =0 \, , 
\label{linear_b} 
\ea
where the recursion relations are valid for $2 \leq \ell \leq N-1$ and can be solved analytically, giving
\be
b_\ell^\sss{\pm} = \frac{1}{2^\ell}\left[ \left(\frac{\mathcal{C}_0 - 
\sqrt{\mathcal{C}_0^2 -4t^2}}{t}\right)^\ell \mathcal{C}_1^\sss{\pm}
+\left(\frac{\mathcal{C}_0 + \sqrt{\mathcal{C}_0^2 -4t^2}}{t}\right)^\ell \mathcal{C}_2^\sss{\pm}
\right] \, , 
\ee
with $\mathcal{C}_0 \equiv V(N-3) + \tfrac{1}{2}\cot\psi_\sss{-} 
+\mathcal{N}_\sss{+} - \l_\textrm{diag}$ and $\mathcal{C}_{1,2}^\sss{\pm}$ constants to be determined. Proceeding to 
the quadratic terms, the ones proportional to $\b^\sss{+} \a^\sss{-}$ and $\b^\sss{-}\a^\sss{+}$ first give the constraints
\be
\csc (\psi_\sss{-}) \, b_1^\sss{+} = \l_{\textrm{nondiag}} = \csc(\psi_\sss{+}) \, b_N^\sss{-} \, .
\ee
These two constraints, combined with the set of relations (\ref{linear_b}) are sufficient to completely 
determine the constants $\mathcal{C}_{1,2}^\sss{\pm}$ and therefore all coefficients $b_\ell^\sss{\pm}$. 
We conclude that the coefficients $b_\ell^\sss{\pm}$ are given by
\ba
&& b_\ell^\sss{+} = -\frac{\sin(\psi_\sss{-} + (\ell-1)2\eta)}{\sin((N-1)2\eta + \psi_\sss{-} +\psi_\sss{+})} \nonumber \\
&& b_\ell^\sss{-} = -\frac{\sin(\psi_\sss{+} + (N-\ell)2\eta)}{\sin((N-1)2\eta + \psi_\sss{-} +\psi_\sss{+})} \, .
\label{coeffs_final}
\ea
Concerning the rest of the 
quadratic terms, after some algebra and splitting the terms with respect to various states, we conclude that the 
coefficients $B_{k\ell}$ satisfy the following relations
\ba
&&
\big(\Xi^\sss{+}_3 +\mathcal{N}_\sss{-} \big) B_{1N} 
+ t (B_{1N-1} + B_{2N}) + b^\sss{+}_N \csc\psi_\sss{-} + b_1^\sss{-} \csc\psi_\sss{+} =0 \cr
&&  \big(\Xi^\sss{+}_2 - \mathcal{N}_\sss{+}  \big) B_{12} + t \, B_{13} + b^\sss{+}_2 \csc\psi_\sss{-} =0 \cr
&& \big(\Xi^\sss{-}_2 + \mathcal{N}_\sss{-}\big) B_{N-1N} 
+ t B_{N-2N} +  b_{N-1}^\sss{-} \csc\psi_\sss{+} = 0 \cr 
&& \big(\Xi^\sss{-}_4  + \mathcal{N}_\sss{-} \big) B_{\ell N} 
+ t(B_{\ell-1 N}+ B_{\ell+1N} + B_{\ell N-1}) + b_\ell^\sss{-} \csc\psi_\sss{+} = 0, \qquad 1<\ell<N-1 \cr
&& \big(\Xi^\sss{+}_4 - \mathcal{N}_\sss{+} \big) B_{1\ell} + 
t (B_{1\ell-1} + B_{1\ell+1} + B_{2\ell}) +  b_\ell^\sss{+} \csc\psi_\sss{-} =0, \qquad 2<\ell<N \cr
&&  \big(\Xi^\sss{-}_3 - \mathcal{N}_\sss{+} \big)  B_{\ell\ell+1} 
+t (B_{\ell-1\ell+1} + B_{\ell\ell+2}) = 0, \qquad\qquad\qquad\qquad 1<\ell<N-1 \cr
&& \big(\Xi^\sss{-}_5  - \mathcal{N}_\sss{+} \big)B_{k\ell} 
+ t(B_{k-1\ell} + B_{k+1\ell} + B_{k\ell-1} +B_{k\ell+1}) =0, \qquad1<k<N-2,~~ \ell>k+1
\nonumber
\ea
where for the sake of readability we have defined
\be
\Xi_q^{\sss{\pm}} \equiv \l_{\textrm{diag}} \pm \tfrac{1}{2}\cot\psi_\sss{-} - V(N-q) \, .
\ee
In principle, having acquired the exact 
expressions for $b_\ell^\sss{\pm}$, the above set of relations provides the expressions of $B_{k\ell}$ as well. 
Since it is hard to solve these relations analytically, one may resort to a numerical analysis for 
a given number of chain sites. However, solving these equations for small numbers of chain length, we were able 
to observe the emerging pattern which governs the coefficients $B_{k\ell}$. In short, we have found that they are 
eventually given by the very simple expressions
\be
B_{k\ell} = \frac{\sin((N-1)2\eta + \psi_\sss{-} + \psi_\sss{+}) }{\sin((N-2)2\eta + \psi_\sss{-} + \psi_\sss{+})} 
\big ( b_{k+1}^\sss{+} \, b_\ell^\sss{-} + b_k^\sss{-} \, b_{\ell-1}^\sss{+} \big ) \, .
\label{Bkl_coefs}
\ee
We have explicitly confirmed that the expressions (\ref{Bkl_coefs}) satisfy all the generic constraints and
relations derived above. In conclusion, the $M=0$ eigenvector of the model is completely determined for 
an arbitrary number of chain sites. 

The decoupling of higher/lower sectors, due to the nilpotency of the Grassmannian parameters, further constraints 
the expressions of the eigenvectors which correspond to excited states. For generic values of $M$ then, we 
propose that the corresponding eigenvector will be given by the following schematic expression
\ba
|\Psi\rangle_{M} & = &  (c_1 + c_2 \, \a^\sss{+} \b^\sss{-} + c_3 \, \b^\sss{+}\a^\sss{-}) |M \rangle 
+ c_4 \, \b^\sss{+} |M+1 \rangle + c_5 \, \b^\sss{-} |M+1 \rangle \cr
 &+ &  c_6 \, \b^\sss{+}\b^\sss{-} |M+2 \rangle + c_7 \, \a^\sss{+} |M-1 \rangle + c_8 \, \a^\sss{-} |M-1 \rangle +
 c_9 \, \a^\sss{+} \a^\sss{-} |M-2 \rangle \, ,
\label{generic_eigenvector}
\ea
where with $|M\rangle$ we denote the states with $M$ particles present, or equivalently the states with $M$ 
spins down in the spin picture, i.e. $M$ excitations from the ground state. The number of states with the 
same $M$ is given by the binomial coefficient 
\be
\begin{pmatrix}
 N \cr M
\end{pmatrix} 
= \frac{N!}{M!(N-M)!} \, ,
\ee
so that the states $|M\rangle$ in (\ref{generic_eigenvector}) are to be understood as collections of states 
spanning the degeneracy space for a particular $M$. In the same spirit, the coefficients $c_i \in \mathbb{C}$ 
appearing also in (\ref{generic_eigenvector}) are to be interpreted as sets of coefficients of the degenerate
states.

\section{Discussion}
In the present work, we have solved the small polaron model with nondiagonal
boundary conditions. 
The eigenvalues of the transfer matrix have been
extracted by using the fusion hierarchy of the transfer matrices and also by
considering the functional relations for particular values of the anisotropy
parameter. 
Starting from the fusion hierarchy of transfer matrices together with its
truncation at particular values of the anisotropy parameter we have formulated
the spectral problem as a functional TQ equation.  The latter has been solved
by means of appropriate deformations needed to account for the nondiagonal
nature of the model.  The eigenvalues have been found to depend on two sets of
Bethe roots, for which coupled Bethe ansatz equations have been presented.

An interesting aspect of the model's solution is that, unlike in the case of
the XXZ model with nondiagonal boundary conditions, no restrictions emerge for
the boundary parameters.  This extra freedom, as well as the remnants of
particle number conservation leading to sectors of the Hilbert space labelled
by the integer $M$ appear to be inherited from the supersymmetric nature of
the model.  Furthermore, supersymmetry heavily restricts the structure of the
eigenvectors and has rendered possible to exactly compute the $M=0$ eigenstate
of the model. A more detailed analysis should provide the complete expressions
of all eigenvectors.

Since supersymmetry lifted any possible constraints between the boundary
parameters for the small polaron, it is interesting to consider other
supersymmetric models with nondiagonal boundaries.  Particularly interesting
would be an extension of the analysis of the supersymmetric t-J model with
open boundaries \cite{Gonz94,Essl96} to this case.  In this case nondiagonal
terms breaking either the $U(1)$ charge symmetry or the $SU(2)$ spin symmetry
of the model can be added.  
The latter problem has been partially solved by Galleas \cite{Gall07} and
leads to problems similar to those encountered in the XXX Heisenberg chain
with non-diagonal boundary fields.  As to boundary terms breaking the charge
symmetry of the model we expect that they can be dealt with in a similar
manner as in the small polaron model.

Another route to follow is to consider operator valued representations of the
reflection algebras \cite{Sklyanin:1988yz, FrSl99, FrPa06} instead of
$c$-number solutions, and attempt to solve the model for boundary conditions
breaking the bulk symmetries.  The construction of suitable, generalized
Jordan-Wigner transformations for the nondiagonal boundary terms would provide
valuable information in this spirit.

\subsection*{Acknowledgments}
We would like to thank A. Doikou for useful discussions on the subject. 
This work has been supported by the Deutsche Forschungsgemeinschaft under
grant no. Fr 737/6.

\appendix

\section{The Jordan-Wigner transformation}
The Jordan-Wigner transformation is a bijective mapping between spin-1/2 operators and fermionic 
creation/annihilation operators in one-dimensional quantum systems. In the following it will be 
shown how the small-polaron model with periodic and diagonal boundary conditions can be mapped onto 
an XXZ-Heisenberg spin chain by virtue of this very transformation.

Let $\opcd{j}$ and $\opc{j}$ denote the creation and annihilation operators of spinless fermions 
on the \mbox{$j$-th} lattice site, subject to the anticommutation relations
\begin{equation}
	\{ \opcd{k}, \, \opc{\ell}\} \equiv \opcd{k}\ \opc{\ell} + \opc{\ell}\, \opcd{k} = \delta_{k\ell}
	\; ,\quad
	\{ \opc{k},\, \opc{\ell} \} = 0
	\quad \text{and} \quad
	\{ \opcd{k}, \, \opcd{\ell} \}  = 0\ .
\end{equation}
The Jordan-Wigner transformation expresses these operators in terms of spin-1/2 raising and lowering operators 
\mbox{$S_j^+$ and $S_j^-$} at corresponding sites $j$ by means of
\begin{equation}
	\opcd{j} = e^{i \, \phi_j}\ S_j^-
	\qquad\text{and}\qquad
	\opc{j} = e^{-i \, \phi_j}\ S_j^+
	\label{eqAP:JWT}
\end{equation}
where the phase $\phi_j$ is given by
\begin{equation}
	\phi_j \equiv \pi \sum_{\ell=1}^{j-1}\, S_\ell^+\, S_\ell^-\ .
\end{equation}
Recall also that $S_j^\pm = S_j^x \pm i \, S_j^y$ 
and $S_j^x$, $S_j^y$ and $S_j^z$ are two-dimensional representations of the $\mathfrak{su}(2)$ algebra 
\begin{equation}
	[S_\ell^\alpha , S_k^\beta] = i \, \delta_{\ell k}\, \varepsilon^{\alpha \beta \gamma}\, S_k^\gamma
	\quad ,\qquad
	\alpha, \beta, \gamma \in \{x,y,z\}\ .
\end{equation}
Several useful relations can be deduced directly from these definitions, in particular
\begin{equation}
 \begin{aligned}
	\opn{j}  \equiv \opcd{j}\, \opc{j}  &= S_j^-\, S_j^+  &, \quad  \opcd{j}\,   \opc{j+1} &= -S_j^-\, S_{j+1}^+ \ ,\\
	\opnq{j} \equiv \opc{j}\,  \opcd{j} &= S_j^+\, S_j^-  &, \quad  \opcd{j+1}\, \opc{j}   &= -S_j^+\, S_{j+1}^- \ .
 \end{aligned}
\end{equation}
According to (\ref{Hbulk}) the bulk part of the small polaron model is determined by the 
Hamiltonian density
\begin{equation}
	H_{j,j+1} = -t\cdot ( \opcd{j+1}\, \opcd{j} + \opcd{j}\, \opc{j+1} )
	            + V\cdot ( \underbrace{[1-\opnq{j+1}]}_{\opn{j+1}}\cdot 
	            \underbrace{[1-\opnq{j}]}_{\opn{j}} + \opnq{j+1}\, \opnq{j} ) \, , 
\end{equation}
which the Jordan-Wigner transformation (\ref{eqAP:JWT}) maps to
\be
	H_{j,j+1} =  2t\cdot ( S_j^x\, S_{j+1}^x + S_j^y\, S_{j+1}^y ) + 2V\cdot S_j^z\, S_{j+1}^z\ ,
\ee
which is precisely the Hamiltonian density of the XXZ spin chain. 

Along the same lines it is easily shown that the diagonal boundary contributions (\ref{H_boundary}) 
to the open small polaron Hamiltonian, i.e.
\begin{equation}
	H^{\text{diag}} = \mathcal{N}_0 ( \opnq{1} - \opn{1} ) + ( \mathcal{N}_+\, \opnq{N} - \mathcal{N}_-\, \opn{N} )
\end{equation}
map to
\begin{equation}
	H^{\text{diag}} = 2 \mathcal{N}_0\, S_1^z + (\mathcal{N}_+ + \mathcal{N}_-) ( S_N^z + \frac{1}{2} )\ . 
\end{equation}
The fact that $\mathcal{N}_+ \neq \mathcal{N}_-$ in the diagonal boundary terms stems from
the supersymmetric nature of the model.

\end{document}